\documentclass[prl, twocolumn]{revtex4}
\usepackage{amssymb}
\usepackage{amsmath}
\usepackage{epsfig}
\usepackage{bm}

\begin{document}

%




\title{Swapping spin currents: interchanging spin and flow directions}                           

\author{Maria B. Lifshits$^{1,2}$ and Michel I. Dyakonov$^{1}$}  

\affiliation{$^{1}$Laboratoire de Physique Th\'eorique et Astroparticules, Universit\'e 
         Montpellier II, CNRS, France\\ $^{2}$ A.F. Ioffe Physico-Technical Institute, 
	 194021, St. Petersburg, Russia}



\begin{abstract}
We introduce a previously unknown spin-related transport phenomenon, consisting in a 
transformation ({\it swapping}) of spin currents, in which the spin direction and the direction of
flow are interchanged. Swapping is due to the spin-orbit interaction in scattering. It 
originates from the correlation between the spin rotation and the scattering angle. This effect is 
more robust than the skew scattering, since it exists already in the first Born approximation. 
Swapping may lead to the spin accumulation with spin polarization perpendicular to the surface, 
unlike what happens in the spin Hall effect.


\vspace{15 pt}

\begin{raggedleft} 
 {``\it After all this is again a problem which falls into the same \\ 
 cathegory of problems related to the orientation of spin." } \\ 
 From the report of our Referee E \\
\end{raggedleft}
\end{abstract}

\maketitle

\thanks{tratata}

Spin-orbit interaction leads to the mutual transformations of spin and charge currents 
\cite{DP_Jetp_71, DP_Physlett_71, DKh_2008}: the charge current generates the transverse spin current 
and {\it vice versa}. The family of these phenomena includes the anomalous Hall effect in ferromagnets 
\cite{Hall_81, Smit_51, KL_54} and the direct and inverse spin Hall effect \cite{DP_Jetp_71, DP_Physlett_71}. 
After the first experimental observations \cite{Kato, Wunderlich}, the spin Hall effect in semiconductors 
and metals attracted much interest and became the subject of numerous experimental and theoretical studies 
\cite{DKh_2008}.Most of the theoretical work is devoted to the intrinsic (not related to electron scattering) 
mechanism of spin-charge coupling \cite{Murakami, Sinitsyn}, which was first proposed by Karplus and Luttinger 
\cite{KL_54}.

In this Letter we introduce another transport phenomenon belonging to the same family. We will show 
that because of spin-orbit interaction, a given spin current induces not only the charge current, 
but also a transverse {\it spin current} with interchanged spin direction and the direction of
flow. We will refer to this transformation as the {\it swapping} of spin currents.

The notion of spin current was introduced in Ref. \cite{DP_Jetp_71}. It is described by a tensor 
$q_{ij}$ where the first index indicates the direction of flow and the second one shows which 
component of spin is flowing. Following \cite{DKh_2008, D_07} we write down the phenomenological 
equations describing the coupling between spin and charge currents, $q_{ij}$ and $q_{i}$ (more
accurately, $\bm q$ is the electron {\it flow} density, related to the electric current density 
$\bm j$ by $\bm q = -\bm j/e$, where $e$ is the elementary charge). We consider an isotropic material with 
inversion symmetry. Then we have \cite{D_07}:
\begin{eqnarray}
q_{i} = q_{i}^{(0)} +\gamma \varepsilon_{ijk} q_{jk}^{(0)},
\label{eq_1}
\\
q_{ij} = q_{ij}^{(0)} -\gamma \varepsilon_{ijk} q_{k}^{(0)}, 				
\label{eq_2}
\end{eqnarray}
where $q_{i}^{(0)}$ and $q_{ij}^{(0)}$ are the primary currents, which may exist in the absence of spin-orbit interaction, $\varepsilon_{ijk}$ is the unit antisymmetric tensor and $\gamma$ is a dimensionless parameter proportional to the strength of spin-orbit interaction.

Pure symmetry considerations allow for additional terms in  Eq. (\ref{eq_2}) proportional to $q_{ji}^{(0)}$ and $\delta_{ij}q_{kk}^{(0)}$, which describe transformations of spin currents.
In the presence of electric field $\bm E$ and spin polarization $\bm P$, this would result in additional contributions to $q_{ij}$ proportional to $E_j P_i$ and $\delta_{ij}({\bm E \cdot \bm P})$, which were  mentioned already in \cite{DP_Jetp_71, DP_Physlett_71}.  
However the physical origin of these contributions was not understood at the time and the effect was not studied ever since.

We will show that the additional terms should always enter in a combination $q_{ji}^{(0)} - \delta_{ij}q_{kk}^{(0)}$ so that Eq.~(\ref{eq_2}) should be modified as:
\begin{eqnarray}
q_{ij} = q_{ij}^{(0)} -\gamma \varepsilon_{ijk} q_{k}^{(0)} 
				+ \varkappa ( q_{ji}^{(0)}-\delta_{ij}q_{kk}^{(0)}),
\label{eq_3}
\end{eqnarray}
with a new dimensionless parameter $\varkappa$ \cite{REF0}. We will also show that the resulting swapping of spin currents originates from the correlation between the scattering direction and spin rotation during collisions.
This effect is more robust than the spin-charge coupling: the swapping constant
$\varkappa$ exists already in the Born approximation, while $\gamma$ appears only beyond this approximation.
\begin{figure}
\epsfxsize=200pt {\epsffile{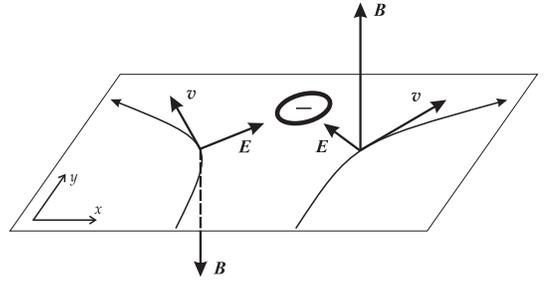}}
\caption
{Schematics of electron scattering by a negative charge. The electron spin
sees a magnetic field ${\bm B}\sim{\bm v} \times \bm{E}$ perpendicular to the trajectory plane.
Note that the magnetic field (and hence the sense of the spin rotation) has opposite directions for electrons scattered to the left and to the right. }
\end{figure}

The physical origin of the swapping effect can be readily  seen from  Fig. 1 illustrating the spin dependent scattering.
The most important element is the magnetic field ${\bm B}$ existing
in the electron's moving frame and seen by the electron spin. 
This field is perpendicular to the plane of the electron trajectory and has opposite signs for electrons moving to the right and to the left of the charged center.
The Zeeman energy of the electron spin in this field is, in fact, the spin-orbit interaction.

Three spin-dependent effects can be seen:

1) The precession of the electron spin around ${\bm B}$ during the collision, leading to the Elliott-Yafet spin relaxation.

2) The spin asymmetry in scattering (the Mott effect, or skew scattering) resulting from the additional force proportional to the gradient of electron Zeeman energy.

3) The {\it correlation} between the directions of electron spin precession and scattering. Indeed, one 
can see that while the spin on the left trajectory is rotated clockwise, the spin on the right 
trajectory is rotated counterclockwise. This correlation and its consequences for spin transport 
were not discussed previously. We will now see that this correlation leads to a transformation of 
spin currents.

Suppose that the incoming electrons move in the $y$-direction and are polarized along $y$ (spin 
current $q_{yy}^{(0)}$). The electrons scattered to the left (right) will acquire a small positive 
(negative) projection of spin on the $x$ axis. This means that the initial $q_{yy}^{(0)}$ spin 
current is partly transformed to $-q_{xx}$. For the case when incoming (along $y$) electrons are 
polarized along $x$, a similar reasoning shows that the initial spin current $q_{yx}^{(0)}$ will 
give rise to $q_{xy}$. Thus in the latter case the spin direction and the direction of flow are
interchanged. Summarizing, one can derive the spin current transformation law: $q_{ji}^{(0)} - \delta_{ij}q_{kk}^{(0)} \rightarrow q_{ij}$, as expressed by Eq. (\ref{eq_3}). 

The reason why the terms $q_{ji}^{(0)}$ and $\delta_{ij}q_{kk}^{(0)}$ enter in such a combination can be explained as follows. Since the spin-orbit interaction is weak, the transformed currents are small compared to the primary currents. Suppose that the primary current contains only one component, say $q_{xx}^{(0)}$. In the first order in spin-orbit interaction there should be no correction to {\it this} component. 
(Similarly, when a vector ${\bm a}$ is rotated on a small angle, the first order correction $\delta {\bm a}$ is perpendicular to ${\bm a}$.) The combination $q_{ji}^{(0)} - \delta_{ij}q_{kk}^{(0)}$ is the only one to meet this requirement.
Generally, the terms  $q_{ji}^{(0)}$ and $- \delta_{ij}q_{kk}^{(0)}$ enter with coefficients whose difference is second order in spin-orbit interaction. 

We now proceed with the formal derivation of Eqs.~(\ref{eq_1}, \ref{eq_3}). Introduce the spin density matrix $\rho_{\alpha \beta}({\bm p})$, where $\alpha$ and $\beta$ are spin indices, ${\bm p}$ is the electron momentum. For a given energy, the evolution of  $\rho_{\alpha \beta}$ due to elastic collisions is described by the following equation \cite{DKh_2008, REF}  
\begin{eqnarray}
	\label{eq_4}
 \!\!\!\!\!\!\!\!	\frac{d\rho_{\alpha \beta}({\bm p})}{dt}  = 
	\!\! \int \!\! d\Omega' \!\!\!\! & & \!\!   \left[
		  { W_{\alpha' \beta'}^{\alpha \beta} \ \rho_{\alpha' \beta'}({\bm p}')  }\right. \nonumber
\\
		-\frac{1}{2}  \!\!\! & & \!\!\!\!	 \left.{\left({
			  W_{\alpha' \alpha'}^{\alpha \beta'} \  \rho_{\beta'\beta }({\bm p})+
			W_{\alpha' \alpha'}^{\beta' \beta } \ \rho_{\alpha\beta'}({\bm p})} \right) }\right],
\end{eqnarray}
 where the integration is done over the angles of ${\bm p}'$ and
\begin{equation}
	\label{W}
	W_{\alpha' \beta'}^{\alpha \beta} = 
	N v F_{\alpha' \bm{p'}}^{\alpha \bm{p}} \left({F_{\beta' \bm{p'}}^{\beta \bm{p}}}\right)^*.
\end{equation}
Here $N$ is the impurity concentration, $\bm{v}=\bm{p}/m$ is the electron velocity, and $F_{\alpha' \bm{p'}}^{\alpha \bm{p}}$ is the scattering amplitude relating the initial state ($\alpha' \bm{p}'$) and the final state ($\alpha \bm{p}$):
\begin{equation}
\label{F}	
	F_{\alpha' \bm{p'}}^{\alpha \bm{p}} = 
		A(\vartheta)\delta_{\alpha\alpha'} + B(\vartheta) \bm{n}\cdot\bm{\sigma}_{\alpha\alpha'},
\end{equation}
$\vartheta$ is a scattering angle, ${\bm \sigma}$ is the Pauli matrix vector and $\bm{n}$ is the unit vector perpendicular to the scattering plane $\bm{n} =\bm{p}'\times\bm{p}/|\bm{p}'\times\bm{p}|$.

The second term in the Eq.~(\ref{F}) originates from the spin-orbit interaction $H_{SO} = \lambda({\bm k}\times \nabla U)\cdot{\bm \sigma}$, where ${\bm k}={\bm p}/\hbar$, $U({\bm r})$ is the scattering potential, and $\lambda$ is the spin-orbit constant.

In the absence of spin-orbit interaction $B(\vartheta)=0$ and 
$W_{\alpha' \beta'}^{\alpha \beta} = 	N v |A(\vartheta)|^2 \delta_{\alpha \alpha'} \delta_{\beta \beta'}$. Then Eq.(\ref{eq_4}) reduces to the conventional Boltzmann equation with the differential cross-section given by $|A(\vartheta)|^2$.
 
Equation (\ref{eq_4}) is applicable when the orbital motion can be considered classically
\cite{semi}. It was previously used to study depolarization during atomic collisions \cite{DP_1978} and the spin-charge coupling for $J=3/2$ holes in the valence band as well as for the carriers in a gapless semiconductor \cite{DKh_1984}, see also \cite{Kh_1984}.

It is convenient to present $\rho_{\alpha\beta}$ in the form
\begin{equation}
\rho_{\alpha\beta}({\bm p}) = \frac{1}{2}
\left[{f({\bm p})\delta_{\alpha \beta}	+ {\bm {\mathcal P}}({\bm p})\cdot\bm{\sigma}_{\alpha\beta}}\right],
\end{equation}
where $f({\bm p}) = {\rm Tr}(\hat{\rho})$ and ${\bm {\mathcal P}}({\bm p}) = {\rm Tr}(\hat{\bm \sigma}\hat{\rho})$
are the particle and spin polarization distributions respectively.

From Eq. (\ref{eq_4}) we obtain a  system of coupled kinetic equations for $f({\bm p})$ and ${\bm {\mathcal P}}(\bm{p})$: 
\begin{eqnarray}
\label{dfdt}
	\frac{d f(\bm{p})}{dt}  = N v
	\!\!\int \!\! d\Omega' \! & [\sigma_1 \!(f({\bm p}')-f({\bm p})) + 
	 \sigma_2 {\bm n}\cdot{\bm {\mathcal P}}(\bm{p}')],
\\
\label{dPdt}
	\frac{d {\bm {\mathcal P}}({\bm p})}{dt}  = N v
	\!\!\int \!\! d\Omega' \!\!\!\!\! & 
	[
	 \sigma_1\!({\bm {\mathcal P}}({\bm p}')-{\bm {\mathcal P}}({\bm p})) +  \sigma_2 {\bm n} f({\bm p}')  \nonumber\\
	 + \  \sigma_3\!\!\!\!\! & {\bm n} \!\times\! {\bm {\mathcal P}}(\bm{p}')+ 
	 					\sigma_4 {\bm n} \!\times\! ({\bm n} \!\times\! {\bm {\mathcal P}}(\bm{p}'))
	],
\end{eqnarray}
where we have introduced four scattering cross-sections
\begin{eqnarray}
\label{sigma}
	&\sigma_1(\vartheta)  =  |A|^2 + |B|^2, 	&\sigma_2(\vartheta)  =  2 \text{Re}(AB^*), \nonumber
\\ \label{eq_10}
	&\sigma_3(\vartheta)  =  2 \text{Im}(AB^*), 	  &\sigma_4(\vartheta)  =  2 |B|^2. 
\end{eqnarray}
As we will see, they have the following physical meaning. The usual transport cross-section is determined by $\sigma_1$, containing a correction $|B|^2$, which is second order in spin-orbit interaction. 
The spin-charge coupling, described by the parameter $\gamma$ in Eqs.~(\ref{eq_1}, \ref{eq_3}), is related to  $\sigma_2$ (skew scattering cross-section). The cross-section  $\sigma_3$ is responsible for the swapping of spin currents. Finally, $\sigma_4$ determines the Elliott-Yafet spin-relaxation rate. These cross-sections were, in fact, introduced by Mott and Massey \cite{Mott} to describe the depolarization of an electron beam in a single scattering event.

We note that in the Born approximation the amplitudes $A(\vartheta)$ and $B(\vartheta)$ have a phase difference $\pi/2$. Thus in this approximation  $\sigma_2 = 0$, however the swapping cross-section  $\sigma_3$ is non-zero.

Because of the rotational invariance of Eqs.~(\ref{dfdt},~\ref{dPdt}), the system of these integral equations can be exactly solved for arbitrary initial distributions $f_0({\bm p})$ and ${\bm {\mathcal P}}_0({\bm p})$ \cite{DP_1978, DKh_1984}. We will derive the coupled equations for the first moments of $f$ and ${\bm {\mathcal P}}$ defined as 
\begin{equation}
	      {\mathcal Q}_i    = \!\int \! \frac{d\Omega}{4\pi} \ v_i f({\bm p}), \;\;\;\;  
        Q_{ij} = \!\int \! \frac{d\Omega}{4\pi} \ v_i {\mathcal P}_j({\bm p}).  
\end{equation}
The charge and spin currents are given by
\begin{equation}
	      q_i    = \!\int \! d^3 {\bm p} \  {\mathcal Q}_i(p),  \;\;\;\; 
        q_{ij} = \!\int \! d^3 {\bm p} \  {\mathcal Q}_{ij}(p).  
\end{equation}
We multiply Eqs. (\ref{dfdt}, \ref{dPdt}) by $v_i$ and integrate over the angles of ${\bm p}$.
The integrals over $d\Omega$ in the right-hand sides can be easily evaluated. The result reads \cite{REF2}
\begin{eqnarray}
\tau_p \frac{d{\mathcal Q}_i}{dt} &=& - {\mathcal Q}_i \  + \gamma\varepsilon_{ijk} {\mathcal Q}_{jk},
\label{eq_dq1}
\\
\tau_p \frac{d{\mathcal Q}_{ij}}{dt} &=& - {\mathcal Q}_{ij} - \gamma \varepsilon_{ijk} {\mathcal Q}_{k} +\varkappa({\mathcal Q}_{ji}-\delta_{ij}{\mathcal Q}_{kk}).				
\label{eq_dq2}
\end{eqnarray}
Here the momentum relaxation time $\tau_p = (Nv\sigma_{tr})^{-1}$ is expressed in the usual way through the transport cross-section $\sigma_{tr}$
\begin{equation}
 \sigma_{tr} = \!\int \! d\Omega \ \sigma_1(\vartheta) (1-\cos\vartheta). 
\end{equation}
The parameters $\gamma$, $\varkappa$, and the Elliott-Yafet spin relaxation time $\tau_s$ are given by
\begin{eqnarray}
 \gamma & =& - \frac{1}{2\sigma_{tr}} \!\int \! d\Omega \ \sigma_2(\vartheta) \sin \vartheta, 
\\ \label{eq_17}
 \varkappa & =& - \frac{1}{2\sigma_{tr}} \!\int \! d\Omega \ \sigma_3(\vartheta) \sin \vartheta, 
\\ \label{eq_18}
 \frac{1}{\tau_s} &  =& \frac{2}{3}Nv \!\int  \! d\Omega \ \sigma_4(\vartheta). 
\end{eqnarray}
Introducing the averaged over angles of ${\bm p}$ spin polarization ${\bm P}(\varepsilon)$, in a similar way we obtain  
\begin{equation}
\label{dP}
\frac{d {\bm P}(\varepsilon)}{dt}  = - \frac{{\bm P}(\varepsilon)}{\tau_s}. 
\end{equation}
For a given energy $\varepsilon = p^2/(2m)$, Eqs.~(\ref{eq_dq1}, \ref{eq_dq2}, \ref{dP}) describe the decay of the charge and spin currents and the spin polarization due to elastic collisions. 

We are interested in the stationary solutions of Eqs. (\ref{eq_dq1}, \ref{eq_dq2}) in the presence of external driving forces, for example in the presence of an electric field. 
The right-hand sides of these equations will acquire additional terms 
${\mathcal Q}_{i}^{(0)}$ and ${\mathcal Q}_{ij}^{(0)}$, which are proportional to the derivative $df_0/d\varepsilon$, where   
$f_0(\varepsilon)$ is the equilibrium distribution function.
For degenerate electrons, $df_0/d\varepsilon \propto \delta(\varepsilon - E_F)$, where $E_F$ is the Fermi energy.
Integrating the stationary solutions of Eqs. (\ref{eq_dq1}, \ref{eq_dq2}) over ${\bm p}$ we finally derive Eqs. (\ref{eq_1}, \ref{eq_3}), where the parameters $\gamma$ and $\varkappa$ are taken at the Fermi energy. 
We note that Eqs. (\ref{eq_1}, \ref{eq_3}) (but not Eqs. (\ref{eq_dq1}, \ref{eq_dq2})) are valid only in the first order in spin-orbit interaction \cite{REF0}.
For non-degenerate electrons, the expression (\ref{eq_17}) should be modified as
\begin{eqnarray}
        \varkappa = - \frac{N}{2 \left\langle {\tau_p} \right\rangle } 
			\left\langle  v \tau_p^2  \!\int \sigma_3 \sin \vartheta \: d\Omega \right\rangle, 
\end{eqnarray}
where $\langle ... \rangle$ means averaging over the electron energy with the equilibrium distribution function.
 
We will now calculate the swapping parameter $\varkappa$ for the Coulomb scattering in the Born approximation. For a positively charged center the scattering amplitudes are 
\begin{eqnarray}
	A(\vartheta) & = & - \left[{ 2 a_B k^2 \: \sin^2(\vartheta/2)}\right]^{-1},\nonumber\\ 
	\label{eq_21}	
	B(\vartheta) & = & - i \lambda k^2 \sin(\vartheta) A(\vartheta),	
\end{eqnarray}
where $a_B = \hbar^2 \epsilon /me^2$ is the Bohr radius, $\epsilon$, $m$ and $e$ are the dielectric constant, electron effective mass, and charge respectively. Using Eqs. (\ref{eq_10}, \ref{eq_17}), and (\ref{eq_21}) we obtain
\begin{eqnarray}
        \varkappa = 2\lambda k^2. 
\end{eqnarray}
In semiconductors with the band structure of GaAs, in the limit of small effective mass, the Kane model gives \cite{Gantm} $\lambda =\hbar^2 / 4 m E_g $ for $\Delta\gg E_g$ and $\lambda =(\hbar^2 / 3 m E_g)(\Delta/E_g)$ for $\Delta\ll E_g$, where $E_g$ is the band gap and $\Delta$ is the spin-orbit splitting of the valence band. Hence 
\begin{eqnarray}
\label{eq_23}
        \varkappa = \frac{E_F}{E_g} 															&{\rm for} &\Delta\gg E_g, 
\\ 
\label{eq_24}
        \varkappa = \frac{4}{3}\frac{E_F}{E_g}\frac{\Delta}{E_g}  &{\rm for} &\Delta\ll E_g. 
\end{eqnarray}

For a bulk electron concentration of $10^{17}{\rm cm}^{-3}$ we calculate a quite large value 
$\varkappa = 0.3$ for InSb (Eq. (\ref{eq_23})) and $\varkappa = 0.003$ for GaAs (Eq. (\ref{eq_24})).

To observe the swapping effect one should obviously have a primary spin current. Most easily this is achieved by passing 
an electric current through a ferromagnet. However the strong exchange magnetic field will tend to destroy spin currents 
with polarization different from that of the majority spins. Whether the swapping effect in ferromagnets can produce a 
measurable effect, or not, is an interesting question, which we do not address here.
 
\begin{figure}
\epsfxsize=220pt {\epsffile{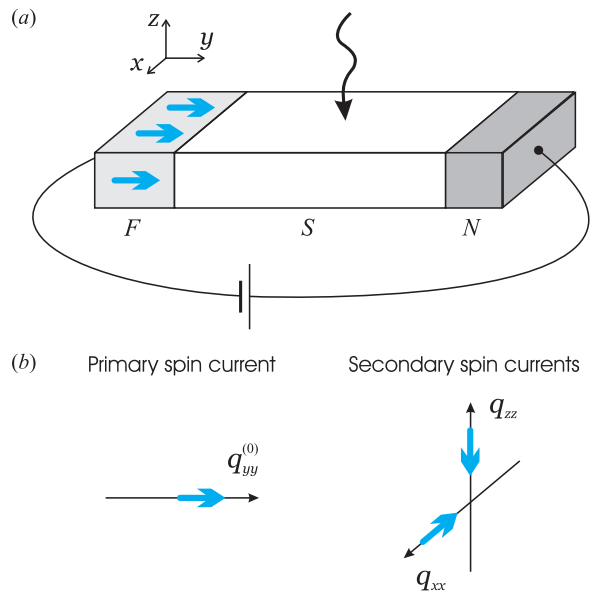}}
\caption{
Schematics of a proposed experiment revealing the swapping of spin currents.
{\it a} - the $q_{yy}^{(0)}$ spin current is electrically injected in a semiconductor ({\it S}) through  a ferromagnetic contact (\textit{F}) with a magnetization along the $y$ axis. 
The wavy arrow symbolizes optical detection of the $z-$component of spin polarization near the surface.
{\it b} - the swapping effect transforms the primary spin current $q_{yy}^{(0)}$ into $q_{xx}$ and $q_{zz}$.
This should lead to the appearance of excess spin polarization at the lateral boundaries of the sample ($P_z<0$ at the top face and $P_z>0$ on the bottom one).
This polarization may be detected optically.
}
\label{fig_2}
\end{figure}

A possible way to see the swapping effect in semiconductors, where the exchange field is negligible, is presented in Fig. \ref{fig_2}. The primary spin current $q_{yy}^{(0)}$ is produced by spin injection in a semiconductor 3D sample through a ferromagnetic contact. Swapping will result in the appearance of transverse spin currents $q_{xx} = q_{zz} = - \varkappa q_{yy}^{(0)}$.
Those secondary currents will lead to an excess spin polarization near the lateral boundaries of the semiconductor sample, which could be detected by optical means. At the top face there will be a polarization $P_z<0$ and at the bottom face $P_z>0$. (Similarly $P_x<0$ at the front face and $P_x>0$ at the back face.)
The accumulation of spins polarized perpendicular to the surface distinguishes this manifestation of swapping from the spin Hall effect, where the accumulated spins are parallel to the surface.

It seems plausible that not only the scattering mechanism considered here, but {\it any} mechanism leading to
the spin Hall effect, will also give rize to the swapping phenomenon. However for the moment we do not see how
this could happen due to the ``intrinsic" mechanism \cite{Murakami, Sinitsyn}. This problem should be addressed in future work. 

In summary, we have considered a new phenomenon caused by spin-orbit interaction and consisting in swapping of spin currents. 
This effect originates from the correlation between the spin rotation and the direction of scattering.

This work was supported by RFBR, the
programs of the Russian Ministry of Education and Science, and the
President Grant for young scientists (MD-1717.2009.2).

\section {APPENDIX. Report of Referee B}

The authors claim to have found a new ``spin current swapping effect"
based mostly on symmetry considerations and obtaining a result
of polarization at the edges due to a polarized charge current in
linear order to the spin-orbit coupling.

I find the result problematic from several considerations (see
below), with some of the explanations and some of the partial
conclusions already present in different forms in other publications
(although the authors seems to be, according to the authors, the
only active spintronic researchers in the field), and therefore
cannot recommend the manuscript for publication in Physical Review
or Physical Review Letters.

The authors, although not specified in the abstract, restrict
their study to the case of weak spin-orbit coupling (otherwise
the notion of spin-current and their symmetries considerations
do no longer hold) and identify several tensor components in the
equations relating spin current and charge current and identify a
relating term Kappa which they try to calculate.

They make a qualitative argument for the effect based on the
previously used explanations of the work of Sinova et al 04 (not
cited here) which identify the rotation from effects in between
collisions. Note also that similar couplings as the ones mentioned
have been obtained in the linear response theory of other authors
(e.g. Sinitsyn et al PRB 081212 (04)); also, many of their results
I believe to be nothing more than the Edelstein effect which has
the same symmetry (but again, not cited by the authors).

The authors follow the procedures of Ref. 11 (which is mostly
equivalent to the result of Ref. 1, simply reprinted 30 years later
in Physical Review Letters) and seem to derive (with many assumptions
along the way) Eq. 20 which they obtain the final result Eq. 22 which
they proceed to estimate for GaAs and InSb. One obvious problem
is that their result has not been observed the experiments (ref. 7)
and according to their geometry it should have been observed but of
course it was not, suggesting strongly that in fact $\varkappa=0$. Such
spin-accumulation would have been observed, very easily according
to their estimates, in these experiments. A linear result in SO
coupling in these case should be highly suspicious since it should
have been extremely evident a long time ago.

I also find it intriguing that the authors suggest that experiments
7 and 8 generated a large interest and debate when in fact in 2004 a
very large debate was ignited by prior theory works, one of which the
authors use similar diagrams (Fig. 1) in the heuristic explanations
of their argued effect.

Given all the inconsistencies that I find in their argument I
cannot agree at this point to recommend the submitted manuscript
to Physical Review and much less to Physical Review Letters.


\begin{thebibliography}{}
\bibitem{REF0}
Equations~(\ref{eq_1}-\ref{eq_3}) are valid in the first order in spin-orbit interaction. The exact equations can be obtained from Eqs.~(\ref{eq_1}-\ref{eq_3}) by replacing $q_i^{(0)}\rightarrow q_i$ and $q_{ij}^{(0)}\rightarrow q_{ij}$ in the terms containing $\gamma$ and $\varkappa$.
\bibitem{REF} 
This equation can be understood by considering the change of the density matrix $\hat \rho$ resulting from a single collision $\delta \hat \rho = \hat S \hat \rho \hat S^\dag -\hat \rho$, where $\hat S = \hat 1 - i \hat T$ is the scattering matrix. 
Using the unitarity of $\hat S$, one finds for  $\delta \rho_{ab}$ an expression similar to the integrand of Eq. (\ref{eq_4}) with $W^{ab}_{a'b'} = T^a_{a'}(T^b_{b'})^*$.
Summing over all collisions per unit of time and taking into account that $T^a_{a'}$ is proportional to the scattering amplitude one obtains Eq. (\ref{eq_4}). Note that in our case the density matrix $\hat \rho$ is diagonal in the momentum space.
\bibitem{semi} Such an approach is often called "semiclassical": everything is classical, except
spin.
\bibitem{REF2} In the right-hand side of Eq. (\ref{eq_dq2}) we have omitted an additional term $\delta {\mathcal Q}_{ij}$ which is second order in spin-orbit interaction, where $\delta = - 2/(3\sigma_{tr})\int  \! d\Omega \ \sigma_4(\vartheta) \cos\vartheta$.
\end{thebibliography}
\end{document}